\documentclass[11pt]{article}
\usepackage{amsmath,amssymb}
\usepackage{setspace,subfig}
\usepackage{natbib}
\usepackage{relsize}
\renewcommand{\harvardurl}[1]{\textbf{URL:} \url{#1}}
\usepackage{booktabs}
\usepackage{tkz-graph}
\usepackage{pgf}
\usepackage{setspace}
\usepackage{tikz}
\usepackage{times}
\usetikzlibrary{arrows,shapes.arrows,shapes.geometric,
	shapes.multipart,backgrounds,decorations.pathmorphing,positioning,fit,automata}
\usepackage[latin1]{inputenc}
\usepackage{soul,color}
\usepackage{chngcntr}
\usepackage{amsthm}
\usepackage{verbatim}
\usepackage{enumitem}
\usepackage{changepage}
\usepackage{amssymb}

\newcommand{\bm}[1]{\mbox{\boldmath{$#1$}}}

\newcommand{\h}{\mathcal{H}}
\usepackage{caption} 
\usepackage{xfrac}
\usepackage{graphicx}
\newcommand{\ind}{\rotatebox[origin=c]{90}{$\models$}}

\newtheoremstyle{note}
{8pt}
{8pt}
{}
{}
{\bfseries}
{:}
{.5em}
{}

\theoremstyle{note}
\newtheorem{theorem}{Theorem}
\newtheorem{lemma}{Lemma}

\newtheorem{remark}{Remark}
\usepackage[T1]{fontenc}
\newtheorem{assumption}{Assumption}
\newtheorem{definition}{Definition}

\allowdisplaybreaks
\newcommand\numberthis{\addtocounter{equation}{1}\tag{\theequation}}

\usepackage[left=1in,right=1in,top=1in,bottom=1in]{geometry}
\usepackage{algorithm}
\date{}
\usepackage{tikz}

\usepackage{soulpos}
\soulregister\cite7
\soulregister\citep7
\soulregister\ref7
\soulregister\emph7
\soulregister\eqref7
\soulregister\pageref7
\definecolor{mygreen}{RGB}{144,241,47}

\newcommand{\mG}{\mathcal{G}}
\newcommand{\ug}{\underline{g}}
\newcommand{\bg}{\bm g}

\newcommand{\upg}{\bar{g}}

\newcommand{\odd}{observed data distribution}
\newcommand{\sir}{feasible region}
\newcommand{\supple}{the Supplementary Materials}

\newcommand{\diff}{\Delta(z_1,z_2;g)}
\newcommand{\hnc}{{\mathcal{H}_{0,c}}}
\newcommand{\inda}{\textcolor{white}{\quad \quad }  }	
\def\bSig\mathbf{\Sigma}

\usepackage[figuresright]{rotating}

\usepackage{authblk}
\author[1]{Linbo Wang\thanks{Address for correspondence: Linbo Wang, Department of Biostatistics, University of Washington, Box 357232, Seattle, WA 98195, USA \\
		Email: lbwang@uw.edu}}
\author[1]{Thomas S. Richardson}
\author[1,2]{Xiao-Hua Zhou}
\affil[1]{University of Washington, Seattle, USA}
\affil[2]{Veterans Affairs Puget Sound Health
	Care System, Seattle, USA}

	{\raise0.5ex\hbox{$#1$}\! \left/ \! \lower0.5ex\hbox{$#2$}\right.}
	
\begin{document}

	\title{Causal analysis of ordinal treatments and binary outcomes under truncation by death}

	\clearpage \maketitle
	

\begin{abstract}	
{	It is common that in multiarm randomized trials, the outcome of interest is ``truncated by death,'' meaning that it is only observed or well defined conditioning on an intermediate outcome. In this case, in addition to pairwise contrasts, the joint inference for all treatment arms is also of interest. 	Under a monotonicity assumption we present methods for both pairwise and joint causal analyses of ordinal treatments and binary outcomes in presence of truncation by death. We illustrate via examples the appropriateness of our assumptions in different scientific contexts.
 }
\end{abstract}	

%
%
%
%
%
%

\section{Introduction}
\label{sec:intro}

In multiarm randomized trials, researchers  are often interested in analyzing treatment effects on an outcome that is measured or well defined only when an intermediate outcome takes certain values \citep{robins1986new,rubin2000causal,rubin2006causal,egleston2007causal,chiba2011simple,ding2011identifiability}.  For example, consider a multiarm randomized HIV vaccine trial. Scientists might be interested in evaluating vaccine effects on HIV viral load as it correlates with infectiousness and disease progression \citep{hudgens2003analysis,gilbert2003sensitivity}. However, HIV viral load is typically measured only for infected individuals.  Two problems occur in this case: in general, there are many  potential comparisons that can be made between different vaccination groups among infected subjects; moreover, these comparisons are subject to selection bias as the vaccine may affect susceptibility to HIV infection.
In the simple case of a two-arm trial, to deal with the selection bias problem, several authors have proposed to  consider  the vaccine effects on viral load among the always-infected stratum, the subpopulation who would become infected regardless of whether they are vaccinated or not \citep[e.g.,][]{hudgens2003analysis,gilbert2003sensitivity}.   
{ However,  there has not been much work on analyzing this type of trial with more than two arms.  }

	By convention, the intermediate outcome  is called ``survival,'' and we say the final outcome is ``truncated by death'' if it is only observed and/or well-defined for ``survivors.'' Thus in the HIV vaccine example above, the always-infected stratum is referred to as the ``always-survivor'' stratum.  The causal contrast among the always-infected subjects is hence called the (always-)survivor average causal effect (SACE) \citep[][\S 12.2]{rubin2000causal,robins1986new}.
	
	In general, even in a two-arm trial, the SACE is not identifiable without strong untestable assumptions. As a result,  there are no consistent tests for detecting non-null vaccine effects in the always-infected stratum.  Instead, under some reasonable assumptions, \cite{hudgens2003analysis} tested the null hypothesis presuming the maximal degree of selection bias. 
Their approach is  related to estimation of bounds on SACE, which has been extensively studied in literature. For example, \cite{zhang2003estimation} developed  bounds on SACE under various assumptions including the monotonicity assumption and the stochastic dominance assumption. \cite{imai2008sharp} provided an alternative proof that the bounds of \cite{zhang2003estimation} are sharp by formulating the truncation-by-death problem as a ``contaminated data'' problem.  
These testing and estimation methods are appealing in practice as they don't rely on strong identifiability assumptions.
	
{	However, so far as we are aware, there has not been much discussion on testing and estimation of SACEs in a multiarm trial, which is fairly common in medical practice {\citep{schulz2005multiplicity}}.  Prior to our work, {\cite{lee2010causal}} considered a sensitivity analysis approach to identify all SACEs in a three-arm trial. Their identification results rely on a strong parametric assumption and several sensitivity parameters. In this article, we instead propose a framework to systematically analyse SACEs in a general multiarm trial without strong identification assumptions. To the best of our knowledge, our method is also the first that is readily applicable to   randomized trials with more than three treatment arms under truncation by death.}
	
The testing and estimation of SACEs in a multiarm trial are more challenging compared to two-arm trials.  Firstly,  in general there are many different SACEs that are well-defined.    As we show later in Section {\ref{sec:causal_estimands}}, consideration of all SACEs (as in {\cite{lee2010causal}}) can lead to paradoxical non-transitive conclusions. Hence we instead restrict our attention to comparisons within the ``finest'' (principal) strata, thereby avoiding this difficulty. Secondly, one needs to distinguish between an overall analysis of treatment effects and a separate analysis of each individual contrast. In the simple setting without truncation by death, it is widely known that compared to all pairwise comparisons with correction for multiple comparisons, an overall analysis such as an ANOVA test 
often provides more power for testing the overall treatment effect in a multiarm trial.  
 When truncation by death is present, because of the non-identifiability of SACEs, this advantage becomes more fundamental as non-identifiability remains even when the sample size goes to infinity. 
	In contrast to {\cite{lee2010causal}}, we distinguish between simultaneous versus marginal inference for SACEs, and argue that they should be used to answer different questions. In particular, we show that compared to marginal inference procedures, our proposed simultaneous inference procedures provide more power for testing the overall treatment effect and the advantage remains even with an infinite sample size. Thirdly, the simultaneous inference problem is unique to a multiarm trial. Again, since SACEs are not identifiable, traditional statistical inference tools for multiarm trials without truncation by death are not directly applicable to our setting. Instead, we develop novel simultaneous inference procedures to test an overall treatment effect, and show that they have desirable asymptotic properties. We also generalize the marginal inference procedures for a two-arm trial to get sharp bounds on SACEs for a general multiarm trial. To focus on addressing these challenges, in this paper, we restrict our attention to trials with ordinal treatment groups and binary outcomes. 
	{
	The rest of this paper is organized as follows. In Section {\ref{sec:framework}}, we introduce our notations, assumptions and define our causal estimands. We also address the transitivity issue  and identify three specific testing and estimation questions that may arise in a general multiarm trial with truncation by death. We then propose three novel procedures that answer these  questions in Sections {\ref{sec:test}}, {\ref{sec:bound}} and {\ref{sec:marginal}}. In Section {\ref{sec:test}}, we  discuss the unique challenges for hypothesis testing with non-identifiable parameters, and develop a novel step-down testing procedure to test the overall treatment effect in this situation. In Section {\ref{sec:bound}}, we develop a linear programming algorithm to test an overall clinically relevant treatment effect. In Section {\ref{sec:marginal}}, we derive the sharp marginal bounds for each causal contrast of interest. In Section  {\ref{sec:data}}, we illustrate the  proposed  procedure with  real data analyses. Results from simulation studies can be found in the \supple.  We end  with a discussion in Section {\ref{sec:discussion}}.
}
	
	The programs that were used to analyse  the data can be obtained from \\ {\tt http://wileyonlinelibrary.com/journal/rss-datasets}.
	
	\section{Framework}
	\label{sec:framework}
	
	\subsection{Data structure and assumptions}
	
	Consider a multiarm trial with a control arm and multiple arms of active treatment. Let $Z$ be an ordinal treatment variable, where $Z=0$ corresponds to the control treatment, and $Z \in \{1,\ldots,m\}$ corresponds to different arms of  active treatment. In what follows, we use the terminology ``treatment arms'' and ``treatment levels'' interchangeably.
	We assume that each subject has $m+1$ dichotomous potential outcomes $Y(z),z=0,\ldots,m$, where $Y(z)$ is defined as the outcome that would have been observed if the subject had been assigned to treatment arm $z$.  Similarly, we define $S(z)$ as the potential survival status under treatment assignment $z$. We assume $Y(z)$ is well-defined only if $S(z)=1$. In other words, the outcome of interest is well-defined only for subjects who survive to the follow-up visit. We also assume that the observed data $(Z_i, S_i, Y_i; i=1,\ldots, N)$ are independently drawn from an infinite super-population. 
	
	Let $G=(S(0),\ldots,S(m))$ denotes the \emph{basic principal stratum}  \citep{frangakis2002principal}. If we let the letter $L$ denote $S(z)=1$ (meaning ``live'') and the letter $D$ denote $S(z)=0$ (meaning ``die''), then $G$ can be rewritten as a string consisting of the letters ``L'' and ``D.'' For example, in a three-arm trial, $G_i=DLL$ indicates that subject $i$ would die under control, but would survive under active treatment 1 or 2.
	
	
	We make the following assumptions.
	\begin{assumption} \label{assump:sutva} Stable unit treatment value assumption (SUTVA \citep{rubin1980comment}): there is no interference between units, and there is only one version of treatment.
	\end{assumption}
	Under the SUTVA, the observed outcome equals the potential outcome under the observed treatment arm, namely $Y=Y(Z)$ and $S=S(Z)$.

	\begin{assumption} \label{assump:random} Random treatment assignment:
$
		Z \ind (S(0),\ldots,S(m), Y(0),\ldots,Y(m)).
$
	\end{assumption}
	

	\begin{assumption}\label{assump:monotonicity} {Monotonicity:}
$
	S_i(z_1) \geq S_i(z_2),   i=1,\ldots, N, z_1 \geq z_2.
$
	\end{assumption}
	
	The monotonicity assumption is sometimes plausible in social science studies if the treatment options can be reasonably ordered. For example, in randomized experiments evaluating the effect of incentives on survey response quality,  it is intuitive that higher level of incentives would not hurt survey response rates. This assumption tends to be more controversial in medical studies where $S$ represents survival, in which there are often trade-offs between treatment benefits and side effects. 
	
	The only possible strata under the monotonicity assumption are strata of the form $D\cdots D L\cdots L$.    
	To compress notation, we denote all possible principal  strata as 
	$(D^kL^{m+1-k};  k=0,\ldots, m+1),$ where members of principal stratum $D^k L^{m+1-k}$ would die if assigned to the first $k$ treatment arms but would survive if assigned to the remaining $m+1-k$ treatment arms.  
	
%
%

%

%
%
%


	\subsection{Causal estimands and questions}
		\label{sec:causal_estimands}
		
	For randomized trials with two treatment arms, it is common  to estimate the average causal effect in the $LL$ stratum \citep{kalbfleisch1980statistical, robins1986new,rubin2000causal}, the only subgroup for which both of the potential outcomes are well-defined:
			$SACE = E[Y(1) - Y(0)\mid G=LL].$
	In a general multiarm trial,  researchers may be interested in comparisons of potential outcomes  within the same
	basic principal stratum. For example, in the case where we have three levels of treatment: 0, 1, 2, the target estimands are $E[Y(2)-Y(1)\mid G=LLL], E[Y(1)-Y(0)\mid G=LLL], E[Y(2)-Y(0)\mid G=LLL]$ and $E[Y(2)-Y(1)\mid G=DLL]$.
	These contrasts are causally meaningful as the memberships of  basic principal strata are defined at baseline. 

	{To define the causal estimands for a general multiarm trial, we first introduce some notation. Let  $\mu_g^z \equiv E[Y(z)\mid G=g]$ denote the mean potential outcome under treatment assignment $z$ in basic principal stratum $g$. Also, let
		$\mathcal{M}(g)$  denote  \emph{the minimal treatment level under which members of principal stratum $g$ can survive}.   In other words, for members of principal stratum $g$, $S(z)=1$ if and only if $z \geq \mathcal{M}(g)$. Consequently,
		 $\mu_g^z \text{ is well-defined if and only if } z \geq \mathcal{M}(g).$
Under the monotonicity assumption, all basic principal strata take the form $g=D^k L^{m+1-k}$. By definition, $\mathcal{M}(D^k L^{m+1-k}) = k$.	Also let
		 $\Omega_k = \{g: \mathcal{M}(g) \leq k\}$ denote \emph{the collection of basic principal strata whose members would  survive if assigned to treatment arm $k$}. The pairwise causal   estimands in a multiarm trial then take the form }\begin{equation}\label{comp} \diff \equiv  \mu_g^{z_1} -
	\mu_g^{z_2}, \text{ where } g\in \Omega_{m-1}, z_1>z_2 \geq \mathcal{M}(g).
\end{equation}
		{For notational simplicity, in this article, when we write the notation $\mu_g^z$ and $\Delta(z_1,z_2;g)$, we always assume that it is well-defined.
	We also note that the parameters involved in defining the causal contrasts $\Delta(z_1,z_2;g)$ are contained in the parameter vector $\bm{\mu}_{m-1} \equiv (\mu_g^z; g\in \Omega_{m-1}, z\geq \mathcal{M}(g))$.}
	
Other meaningful causal contrasts  are made within \emph{coarsened principal strata}, defined as groups that combine several basic principal strata \citep{cheng2006bounds}. For example, in the case of a three-arm trial, the contrast $E[Y(2) - Y(1)\mid G\in \{LLL,DLL\}]$ is also causally meaningful as memberships of the coarsened principal strata $\{LLL,DLL\}$ are also defined at baseline. {Some previous researchers} hence consider coarsened principal strata causal effects together with basic principal strata causal effects \citep[e.g. ][]{lee2010causal}. However, as  \cite{robins1986new} noted,   if one were to compare $E[Y(2)-Y(0)\mid G=LLL]$, $E[Y(1)-Y(0)\mid G=LLL]$ and $E[Y(2)-Y(1)\mid G\in \{LLL,DLL\}]$ simultaneously, it is possible that the last two comparisons are both positive while the first one is negative. This lack of transitivity limits the interpretability of causal effects defined within coarsened principal strata. In contrast, transitivity holds if limited to basic principal strata (e.g., $LLL$).  Hence in this article, we are primarily interested in comparisons between potential outcomes in the same  \emph{basic} principal stratum.

	On the other hand, as \cite{Robi:Rotn:Vans:disc:2007} noted, the size of each basic principal stratum is likely to be very small and consequently, each comparison in (\ref{comp}) only applies to a small portion of the population. Hence for randomized trials  with more than three treatment arms, we may have limited power to test treatment effects for each basic principal stratum. What is more, we run into the problem of multiple comparisons as there are multiple treatment arms and multiple basic principal strata. 
	
	Therefore, we first consider
	testing the global null hypothesis that the treatment is not effective in any of the basic principal strata (for which some treatment comparison is well-defined). This question is scientifically relevant.  For example, in a HIV vaccine trial, 
testing the global null addresses whether there exists a mechanism through which the vaccine alters viral load in infected individuals \citep{shepherd2006sensitivity}.  Secondly, clinicians may also be interested in whether the overall treatment effect is clinically meaningful so that the active treatment is promising in clinical practice. For this purpose, an overall treatment effect may be declared only if it is greater than  the clinical margin of relevance specified by clinicians.  
Finally,  besides an overall treatment effect, scientists and clinicians  may also be interested in isolating the non-zero/non-trivial causal contrasts.  
In summary, the following questions are of interest with a multiarm trial:
\begin{enumerate}
	\item Is there evidence of the existence of \emph{non-zero} average treatment effects for at least one basic principal stratum between at least two treatment arms?
	\item Are there \emph{clinically relevant} average treatment effects for at least one basic principal stratum between at least two treatment arms?
	\item {Can we find the specific principal strata and treatment arms that correspond to the overall non-zero/clinically relevant treatment effect, if such exists?}
\end{enumerate}
We address these questions in Section \ref{sec:test}, \ref{sec:bound} and \ref{sec:marginal} respectively. Existing  causal analysis literature in multiarm trials with non-identifiable causal estimands \citep{cheng2006bounds,long2010estimating,lee2010causal} focuses on answering the third question. However, as we  explain  later in Remark \ref{remark:marginal2}, one may be able to answer the first two questions even if there is not enough information to answer the third. Hence it is important to consider all three questions.

	\section{Testing treatment effects in a multiarm trial}
	\label{sec:test}
	
	{To find out if there is an overall non-zero treatment effect,} it is desirable to consider the following testing problem:
	\begin{equation}
	\label{eqn:test}
			\mathcal{H}_0: \diff=0, \forall z_1, z_2, g  \quad vs \quad  \mathcal{H}_a: \exists z_1,z_2, g \ \ s.t. \  \ \diff \neq 0,
	\end{equation}
where $\forall$ means ``for all,'' $\exists$ means ``there exists'' and $s.t.$ means ``such that.''
	{The testing problem {\eqref{eqn:test}} is fundamentally different from (and more difficult than)  a standard testing problem, in which one assumes if the observed data distribution was known, one would also know whether or not the hypothesis is true {\citep{lehmann2006testing}}. The main difficulty here is that $\h_0$ is a statement about non-identifiable parameter vector $\bm{\mu}_{m-1} $. In other words, even if the population probabilities $P(S=1\mid Z=z)$ and $P(Y=1\mid S=1,Z=z)$ were known, we could only ascertain that $\bm{\mu}_{m-1} $ resides in a region, and therefore may not know whether $\h_0$ is true or not. }
	
{	Nevertheless, $\bm{\mu}_{m-1} $ is ``partially identifiable''  in the sense that the observed data distribution can narrow down the range in which $\bm{\mu}_{m-1} $ can possibly lie {\citep{cheng2006bounds}}.
	For example, in a three-arm trial, the domain of  $\bm{\mu}_2$ is $[0,1]^6$. However, if the observed 
	data distribution was known,  the feasible region of $\bm{\mu}_2$ would be a subspace in $[0,1]^6$ subject to the following constraints:}
	\begin{flalign*}
		P(Y=1\mid Z=0,S=1) &= \mu_{LLL}^0,  \\
		P(Y=1\mid Z=1,S=1) &= p_{LLL}^1 \mu_{LLL}^1 + p_{DLL}^1 \mu_{DLL}^1,  \\
		P(Y=1\mid Z=2,S=1) &= p_{LLL}^2 \mu_{LLL}^2 + p_{DLL}^2 \mu_{DLL}^2 + p_{DDL}^2 \mu_{DDL}^2, \numberthis\label{eqn:figure1} 
	\end{flalign*}
	where  $p_g^z \equiv P(G=g\mid Z=z,S=1)$ is identifiable under Assumptions \ref{assump:random} and \ref{assump:monotonicity} (see Lemma 1 in \supple).  Figure \ref{fig:basic} provides a graphical representation of the functional relations described in \eqref{eqn:figure1}.

	
	\tikzstyle{pix}=[circle, draw=black, fill=gray!25]
	
	\tikzstyle{pyxz}=[rectangle, draw=black]
	
	\tikzstyle{gamma0}=[ellipse, draw=black, fill=blue!30]
		\tikzstyle{gamma1}=[ellipse, draw=black, fill=blue!10]
				\tikzstyle{gamma2}=[ellipse, draw=black, fill=blue!1]
	\begin{figure}[!htbp]
		\centering 
	\begin{tikzpicture}[scale=0.2]
	\matrix[row sep=.5in, column sep=.0in, ampersand replacement=\&] {
		\&\&\&	\&\& \node (px) [pix] {$\pi_g$};\&\&\&\&
		\\[-10pt]
		\&\&\&	 \node (px1) [pix] {$p_g^1$};\&\&\&\node (px2) [pix] {$p_g^2$};\&\&
		\\
		\node (yz0) [pyxz]{$E[Y\mid Z=0,S=1]$}; \&\& \&
		\node (yz1) [pyxz]{$E[Y\mid Z=1,S=1]$}; \&\& \&
		\node (yz2) [pyxz]{$E[Y\mid Z=2,S=1]$}; 
		\&\&
		\\
		\node (LLL0) [gamma0] {$\mu_{LLL}^{0}$}; \&\&
		\node (LLL1) [gamma0] {$\mu_{LLL}^{1}$}; \&
		\node (DLL1) [gamma1] {$\mu_{DLL}^{1}$}; \&\&
		\node (LLL2) [gamma0] {$\mu_{LLL}^{2}$}; \&
		\node (DLL2) [gamma1] {$\mu_{DLL}^{2}$}; \&
		\node (DDL2) [gamma2] {$\mu_{DDL}^{2}$};		  
		\\
	};
	\path[->]
	(LLL0) edge (yz0)
	(LLL1) edge (yz1)
	(DLL1) edge (yz1)
	(LLL2) edge (yz2)
	(DLL2) edge (yz2)
	(DDL2) edge (yz2)
	(px) edge (px1)
	(px) edge (px2)
	(px1) edge (yz1)
	(px2) edge (yz2)
	;
	\end{tikzpicture}	
			\caption{ A graph representing the functional dependencies in the causal analysis of a three-arm randomized trial with truncation by death. Rectangular nodes represent observed variables; oval nodes represent unknown parameters, with different shadings corresponding to different principal strata. Under the monotonicity assumption, $p_g^z$  can be identified from observed quantities $P(S=1\mid Z=z)$. }
			\label{fig:basic}
		\end{figure}
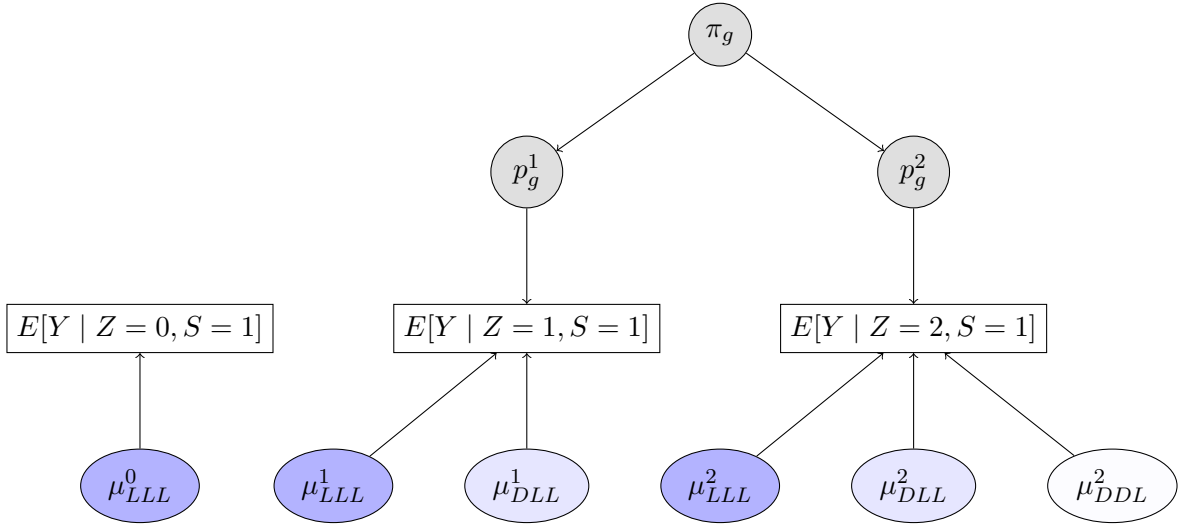
			
{	For a general multiarm trial, if the parameter space defined by $\h_0$ has no intersection with the feasible region of $\bm{\mu}_{m-1} $, one would  know that  $\h_0$ is not true.  In general, we introduce the following notions for hypothesis testing with non-identifiable parameters.}
	
	\begin{definition}
	{	We define a hypothesis relating to  a parameter to  be {\emph{compatible}} with an {\odd}  if the parameter space defined by the hypothesis has a non-empty intersection with the feasible region  of the  parameter under the observed data distribution.}
	\end{definition} 	
In particular,   if a parameter is completely unidentifiable such that the observed data distribution imposes no constraints on the parameter, then all hypotheses relating to that parameter are compatible with the \odd.  On the other hand, if a parameter is identifiable so that its feasible region under the observed data distribution is always a single point set, then all compatible hypotheses are true. 

In general, however, not all compatible hypotheses are true. Nevertheless,  owing to lack of identifiability, a true hypothesis may not be distinguishable from data with an untrue yet compatible hypothesis. This leads to the following notion of sharpness.
	\begin{definition}
		\label{def:sharp}
	We define a test to be {\emph{sharp}} for testing a null hypothesis if when the null is not compatible with the observed data distribution (and is hence untrue), the power of the test tends to 1 when the sample size goes to infinity.
	\end{definition}
	Intuitively, similar to consistent tests, sharp tests are those that maximize power asymptotically. The difference is that as the sample size goes to infinity, with probability tending to 1,  sharp tests reject any hypotheses that are incompatible with the \odd, whereas consistent tests reject any hypotheses they are untrue. In small sample settings, however, the conclusions that one would draw from a sharp test are similar to those from a consistent test. If a hypothesis is rejected, one would conclude that it is untrue (at a certain significance level); if otherwise, no claims about the correctness of the hypothesis would be made. We also note that for a standard hypothesis testing problem as described in {\cite{lehmann2006testing}}, sharp tests are the same as consistent tests.
	 When the null hypothesis concerns  non-identifiable parameters, however,
	 there are in general no consistent tests. Instead, sharpness plays the role of consistency in a standard hypothesis testing problem. 
	 
	The notion of sharp tests is similar in spirit to the notion of sharp bounds, defined as the tightest possible bound given the observed data distribution \citep[e.g.,][]{imai2008sharp}. This notion has also been used implicitly in previous works. For example, \cite{hudgens2003analysis}'s test for SACE in a two-arm trial is sharp.


	Below in Section \ref{sec:step-down}, we develop a sharp test for problem \eqref{eqn:test} under the presumption that the \odd \ is known.  In other words, we assume the sample size is infinite such that there is no stochastic variation in the observed data.  In Section \ref{sec:step_down_bayes} we incorporate sampling uncertainty to our proposed test using a  Bayesian method.

	\subsection{A step-down procedure for testing the global null $\h_0$}
	\label{sec:step-down}
	
 To fix  ideas, we first consider the problem of a three-arm trial, for which $\h_0$ holds if and only if
\begin{equation}
\label{eqn:h0_LLL}
\mu_{LLL}^0 = \mu_{LLL}^1 = \mu_{LLL}^2
\end{equation}
and 
\begin{equation}
\label{eqn:h0_DLL}
\mu_{DLL}^1  = \mu_{DLL}^2.
\end{equation}
We hence propose a two-step procedure. Firstly we test hypothesis \eqref{eqn:h0_LLL}. If \eqref{eqn:h0_LLL} is compatible with the \odd, we then  test if \eqref{eqn:h0_DLL} is compatible with the \odd  \ conditioning on \eqref{eqn:h0_LLL}. 

Specifically,  one can see from Figure \ref{fig:basic}  that   $\mu_{LLL}^0$  is identifiable from the observed data and suppose the feasible regions of $\mu_{LLL}^1$ and $\mu_{LLL}^2$  are $B_{01}$ and $B_{02}$, respectively. If $\mu_{LLL}^0$ is not contained in the intersection of $B_{01}$ and $B_{02}$, then \eqref{eqn:h0_LLL} and hence $\h_0$
are not compatible with the observed data distribution. 
If otherwise, so that \eqref{eqn:h0_LLL} is compatible with the observed data distribution, we then test hypothesis \eqref{eqn:h0_DLL} under the assumption that hypothesis \eqref{eqn:h0_LLL} holds. Note that, under hypothesis \eqref{eqn:h0_LLL},   $\mu_{LLL}^1$ and $\mu_{LLL}^2$ are  identifiable. Consequently,    $\mu_{DLL}^1$  is  identifiable. Suppose the feasible region of $\mu_{DLL}^2$ under the constraint \eqref{eqn:h0_LLL} is  $B_{12}$.   If $\mu_{DLL}^1$ is not contained in $B_{12}$, we conclude that \eqref{eqn:h0_DLL} is not compatible with the \odd  \ under the constraint \eqref{eqn:h0_LLL} and  hence reject  $\h_0$.  If otherwise,  we conclude that $\h_0$ is compatible with the observed data distribution.


	Algorithm \ref{alg:general} generalizes the procedure  described above  to general multiarm trials.   Theorem \ref{thm:algorithm1} states the asymptotic optimality of Algorithm \ref{alg:general}.  The proof  is provided in \supple.

	\begin{algorithm}
		\caption{\ \ A step-down algorithm for testing the global null hypothesis $\h_0$}
		\label{alg:general}
		
		\begin{enumerate}
			\item {\bf Set} $k=0$
			
			\item {\bf For} $z=k,\ldots, m$ 			\vspace{0.2cm} \\
			\textcolor{white}{\quad \quad }  {obtain} the \sir \ (under the maintained assumptions) $B_{kz}$ for  $\mu_{D^k L^{m+1-k}}^z$ (see Theorem \ref{thm:feasible}) 
			\item \quad \vspace*{-0.47cm} \begin{equation*}
				\hspace*{-12.6cm} 	\text{\bf If} \mathop\cap\limits_{z=k,\ldots,m} B_{kz} = \emptyset 
			\end{equation*}
			\textcolor{white}{\quad \quad } reject $\h_0$; report k; stop\\
			\textcolor{white}{\quad \quad } {\bf else} \vspace*{-10pt}  \\
		\textcolor{white}{\quad \quad } 
			 \begin{equation} 
			\label{eqn:assump_working} \hspace*{-5.6cm} 	\text{set }
				\mu_{D^k L^{m+1-k}}^k = \cdots = \mu_{D^k L^{m+1-k}}^{m}
			\end{equation}
			\item {\bf If} $k = m$ \vspace*{2pt}\\
			\textcolor{white}{\quad \quad } fail to reject $\h_0$ and stop \\
			\textcolor{white}{\quad \quad }{\bf else} \vspace*{-0pt}  \\\textcolor{white}{\quad \quad } set $k = k+1$ and go to Step 2
		\end{enumerate}
	\end{algorithm}
	\begin{theorem}
		\label{thm:algorithm1}
		The test given by Algorithm \ref{alg:general} is sharp for testing $\h_0$. In other words, it is asymptotically optimal for testing $\h_0$ as it maximizes power given the \odd.
	\end{theorem}

	To derive  the feasible regions $(B_{kz}; k=0,\ldots,m, z=k,\ldots,m)$ in Algorithm \ref{alg:general}, we  introduce notation building on \cite{horowitz1995identification}. 
{Let $Q_\mG^z(\cdot)$ denote the distribution (function) of outcome $Y$ among members of subgroup $\mG$ who receive treatment $z$, and $\delta_x(\cdot)$ be a degenerate distribution function localized at  $x$.  As $Y$ is binary, $Q_\mG^z(\cdot)$  is a Bernoulli distribution with mean $m_\mG(z)$:}
$ Q_\mG^z(\cdot) = (1-m_\mG(z))\delta_0(\cdot) + m_\mG(z) \delta_1(\cdot).$
To compress notation, we write $Q_\mG^z(\cdot)$ as $Q_\mG^z$.
Also  let \mbox{$L_{\lambda}(Q)$} and \mbox{$U_{\lambda}(Q)$} be functionals that map a distribution function  $Q$ to the corresponding distributions truncated  at the lower \mbox{$\lambda$} quantile and upper \mbox{$\lambda$} quantile, respectively. 
	Theorem \ref{thm:feasible}  gives the formula for feasible region $B_{lz}$.

	\begin{theorem}
		\label{thm:feasible}
Suppose that the \odd \ is known and \eqref{eqn:assump_working} holds for all $k < l$.  Let $g=D^l L^{m+1-l}$ and $\bg = \mathop\cup\limits_{\upg\in \Omega_z \setminus \Omega_{l-1}} \upg$ be the coarsened principal stratum whose members would survive if assigned to treatment $z$ but would die if assigned to treatment $l-1$. The feasible region of $\mu_g^z$ is  \begin{equation}
		\label{eqn:sir_mugz} B_{lz} = \left[\int y d L_{\omega_{g}^z}(Q_{\bg}^z),         \int y d U_{\omega_g^z}(Q_{\bg}^z)\right],
		\end{equation}
		where $\omega_{g}^z \equiv P[G=g\mid G \in \Omega_z \setminus \Omega_{l-1}] = \left.{p_g^z} \middle/ { \left(\sum\limits_{\overline{g} \in \Omega_z \setminus \Omega_{l-1}} p_{\overline{g}}^z \right)}\right. $ and 		
		$Q_{\bg}^z$ is a Bernoulli distribution with mean 
		$$
		m_{\bg}(z) = \left. {\left(m(z)  -  \sum\limits_{\ug \in \Omega_{l-1}} p_{\ug}^z \mu_{\ug}^z \right)}\middle/ {\left(1-   \sum\limits_{\ug \in \Omega_{l-1}} p_{\ug}^z  \right)} \right.,
		$$
		in which $m(z) \equiv P[Y=1\mid Z=z,S=1]$.
	\end{theorem}

		Intuitively, the bounds of $B_{lz}$ are obtained	by assigning the smallest/largest $\omega_g^z$ portion of observed outcome values  in distribution $Q_{\bg}^z$ to  principal stratum $g$. The proof is in \supple.

		

			\begin{remark}
				Algorithm \ref{alg:general} is a ``step-down'' procedure in the sense that the hypothesis $\h_0$ is decomposed into a series of hypotheses where the first hypothesis concerns 
the first stratum $L^{m+1}$, the second hypothesis concerns 
the second stratum $DL^m$ conditioning on the first hypothesis, and so on. 
			\end{remark}

\subsection{Bayesian procedures}
\label{sec:step_down_bayes}

We have so far developed a sharp test for  problem \eqref{eqn:test}. In practice, however, sampling uncertainty must be taken into account when making statistical inference. Here  we introduce a Bayesian procedure to estimate the posterior probability that $\h_0$  is not compatible with the \odd.  The Bayesian method produces multiple samples of the posterior distribution, thereby reflecting randomness in observed data. 
	
	Let $p(s,y\mid z) = P(S=s,Y=y\mid Z=z)$ and  $p(\cdot,\cdot\mid z) = (p(1,1\mid z), p(1,0\mid z),p(0,\uparrow\mid z))$, where $\uparrow$ indicates that $Y$ is undefined when $S=0$. Define $\bm p = (p(\cdot,\cdot\mid 0), \ldots, p(\cdot,\cdot\mid m))$.
	Under independent Dirichlet priors over the \emph{observed distributions} $p(\cdot, \cdot\mid z),z=0,\ldots,m$, it is easy to sample from the posterior distribution via conjugacy. 
We propose to use Algorithm \ref{alg:bayesian} to calculate the posterior probability that $\h_0$ is not compatible with the \odd.
	
	\begin{algorithm}
		\caption{\ \ A Bayesian procedure for testing $\h_0$}
		\label{alg:bayesian}
		\begin{enumerate}
			\item Place an independent Dirichlet prior $Dir(\alpha_{3z+1},\alpha_{3z+2},\alpha_{3z+3})$ on $p(\cdot,\cdot\mid z)$, $z=0,\ldots,m$. 
			\item Simulate samples $\bm p^{(1)}, \cdots,\bm p^{(M)}$ from the posterior distributions, which are independent Dirichlet distributions
			$$
			Dir(\alpha_{3z+1}+n_{3z+1},\alpha_{3z+2}+n_{3z+2},
			\alpha_{3z+3}+n_{3z+3}), z=0,\ldots,m,
			$$
			where $n_{3z+1} = \sum\limits_{i=1}^N I(S_i=1,Y_i=1, Z_i=z), n_{3z+2} = \sum\limits_{i=1}^N I(S_i=1,Y_i=0, Z_i=z),
			n_{3z+3} = \sum\limits_{i=1}^N I(S_i=0, Z_i=z).$
			\item Run Algorithm \ref{alg:general} with each of the posterior samples satisfying the following inequalities:
			\begin{equation}
			\label{eqn:monotone}
			P(S=1\mid Z=m) \geq \cdots \geq P(S=1\mid Z=1) \geq  P(S=1\mid Z=0)
			\end{equation}
		Note	\eqref{eqn:monotone} characterizes the set of observed data distributions arising from the potential outcome model defined by Assumptions \ref{assump:sutva} - \ref{assump:monotonicity}.
			\item Report the proportion of posterior samples with which  $\h_0$ is rejected. 
		\end{enumerate}
	\end{algorithm}

%
	
	\begin{remark}
		The step-down procedure in Algorithm \ref{alg:general}  has a similar structure to the sequential tests for nested hypotheses discussed by \cite{rosenbaum2008testing}.
His procedure has attractive Frequentist properties since it controls the type I error rate without resorting to multiplicity adjustment.  However, with his methods one proceeds to the next step if the current hypothesis is rejected whereas in our proposal, one proceeds if the current hypothesis is \emph{not} rejected. Moreover, in his context, the parameters of interest are identifiable. Hence Rosenbaum's results are not directly applicable to our case.
	\end{remark}

	\section{Testing clinically relevant treatment effects in a multiarm trial}
	\label{sec:bound}
	
	If a non-zero treatment effect is found using Algorithm  \ref{alg:bayesian}, a natural question arises as to whether the treatment effect is clinically meaningful.  Suppose the margin of clinical relevance is $\Delta_0$ such that a treatment effect smaller than this would not matter in practice, and also suppose that the treamtent effect is clinically meaningful only if a higher treatment level corresponds to a higher mean potential outcome. It is desirable to consider the following testing problem: 
	\begin{equation}
	\label{eqn:test_delta}
	\mathcal{H}_{0,c}: \diff \leq \Delta_0, \forall g, z_1\geq z_2  \quad vs \quad  \mathcal{H}_{a,c}: \exists g, z_1 \geq z_2 \ \ s.t. \ \ \diff >\Delta_0,
	\end{equation}	
	where the letter ``$c$'' in $\mathcal{H}_{0,c}$ is short for ``clinical relevance.''  
{Similar to \eqref{eqn:test}, \eqref{eqn:test_delta} is a testing problem on non-identifiable parameters. However, as the null parameter space is a non-degenerate region in the domain of $\bm{\mu}_{m-1} $, the step-down procedure developed in Section \ref{sec:test} is not applicable. Instead, we define
%
$\Delta_{max}$ to  be the largest $\Delta(z_1,z_2;g)$ that appears in $\h_{0,c}$:
	$\Delta_{max} = \max\limits_{g, z_1 \geq z_2} \Delta(z_1,z_2;g).$
	 \eqref{eqn:test_delta} can then be rewritten in an \emph{equivalent} form using $\Delta_{max}$:
$
	\mathcal{H}_{0,c}: \Delta_{max} \leq \Delta_0  \quad vs \quad  \mathcal{H}_{\alpha,c}: \Delta_{max} > \Delta_0.
	$
The following lemma says the testing problem \eqref{eqn:test_delta} can be translated into the identification problem on $\Delta_{max}$.

\begin{lemma}
	Suppose the sharp (large sample) lower bound for $\Delta_{max}$ is $\Delta_{max,slb}$. A sharp test  would reject $\h_{0,c}$ if and only if $\Delta_{max,slb} > \Delta_0$. 
\end{lemma}

 As $\Delta_{max}$ is a function of $\bm{\mu}_{m-1} $, in general, identifying $\Delta_{max,slb}$ involves minimizing $\Delta_{max}$ subject to the constraints on $\bm{\mu}_{m-1} $ imposed by the {\odd}.  Theorem \ref{thm:feasible_mu_m-1} below says that 
 the feasible region of $\bm{\mu}_{m-1} $ is a convex polytope, defined as an intersection of finitely many half spaces.   Consequently, this optimization problem can be translated into a linear programming problem and  efficiently solved with off-the-shelf software.  See Algorithm 1 in \supple \ for more details.
\begin{theorem}
	\label{thm:feasible_mu_m-1}
	Given the \odd, the feasible region of $\bm{\mu}_{m-1} $ is a subspace in $[0,1]^{dim(\bm{\mu}_{m-1} )}$ subject to the following constraints:
	\begin{flalign*}
	\sum\limits_{g\in \Omega_z} p_{g}^z  \mu_{g}^z  &= m(z) , z = 0,\ldots,m-1; \\
	\max\left(0, m(z) -  p_{D^m L}^z \right) &\leq \sum\limits_{g\in \Omega_{m-1}} p_{g}^z \mu_g^z \leq \min \left(  1  -   p_{D^m L}^z,      m(z)    \right),  z =  m,       
	\end{flalign*}
where	$p_g^z$ is identifiable from data under Assumptions \ref{assump:random} and \ref{assump:monotonicity} (see Lemma 1 in \supple).
	In particular,  the feasible region of $\bm{\mu}_{m-1} $ is a convex polytope.
\end{theorem}

%

	
%
		
		To incorporate statistical uncertainty, one can use Bayesian analysis methods to derive a credible interval for $\Delta_{max,slb}$. 
Specifically, one runs Steps 1-4 in Algorithm \ref{alg:bayesian} to get multiple posterior samples that satisfy the constraint \eqref{eqn:monotone}, and then produces a percentile based credible interval for $\Delta_{max,slb}$ based on the posterior samples. One may also  estimate the posterior probability of rejecting  $\h_{0,c}$ for any given positive value $\Delta_0$ with these posterior sample draws.

	\section{Marginal credible intervals for a given contrast}
	\label{sec:marginal}
	
	If a clinically non-trivial treatment effect is found, 
then it is desirable to identify the principal strata and treatment arms that correspond to this treatment effect.  {In this case, the marginal feasible regions and associated credible intervals for $\Delta(z_1,z_2;g)$ are of interest. }

	If the observed data distribution was known, then the \sir \ for $\Delta(z_1,z_2;g)$ can be obtained from the \sir s for $\mu_g^{z_1}$ and $\mu_g^{z_2}$. Specifically, we have the following theorem.
	
	\begin{theorem}
		Suppose the \odd \ is known, and $B_{\mathcal{M}(g),z_1}$ and $B_{\mathcal{M}(g),z_2}$ are \sir s for $\mu_g^{z_1}$ and $\mu_g^{z_2}$, respectively. Then we have the following results.
		\begin{enumerate}
			\item For $z=z_1, z_2$, $ B_{\mathcal{M}(g),z} = \left[\int y d L_{p_{g}^z}(Q^z), \int y d U_{p_{g}^z}(Q^z)\right]. $
			\item The \sir \ of $\Delta(z_1,z_2;g)$  is 
			$
					\left[\int y d L_{p_{g}^{z_1}}(Q^{z_1}) - \int y d U_{p_{g}^{z_2}}(Q^{z_2}), \int y d U_{p_{g}^{z_1}}(Q^{z_1}) - \int y d L_{p_{g}^{z_2}}(Q^{z_2})\right].
			$
		\end{enumerate}
	\end{theorem}

In practice, credible intervals for $\Delta(z_1,z_2;g)$ can be constructed from posterior sample draws $\bm p^{(1)}, \ldots, \bm p^{(M)}$. These posterior draws may also be used to estimate the posterior probability of rejecting the null hypothesis 
$
		\mathcal{H}_{0,m}: \Delta(z_1,z_2;g) \leq \Delta_0,
$
where the letter ``m'' in $\mathcal{H}_{0,m}$ is short for ``marginal.''

	\begin{remark}
		\label{remark:marginal2}
%

	We remark that even if the observed data provide evidence for the existence of non-zero/non-trivial treatment effects, it is possible that they do not contain information on the specific principal strata and treatment arms that correspond to these treatment effects. {Moreover, unlike the case for multiarm trials without truncation by death, this can happen even  with an infinite sample size.}
	
	We illustrate our point with the following numerical example.  Consider a three-arm trial such that $
	 \pi_{LLL}=\pi_{DLL}=\pi_{DDL}=0.3, \pi_{DDD}=0.1, m(0)=0.3, m(1) = 0, m(2)=0.5,
	 $
	 where $\pi_g \equiv P(G=g)$. 
	 In this case, $\mu_{LLL}^0=0.3$ and $\mu_{LLL}^1 = \mu_{DLL}^1=0$. It follows that $
	 \Delta_{max}
	 =  \max(0, \mu_{LLL}^2-\mu_{LLL}^1,  \mu_{DLL}^2 - \mu_{DLL}^1).
	 $
	 We assume that the sample size is infinite so that we know the \odd.
	 Figure \ref{fig:example2} shows the joint \sir \ of $(\mu_{LLL}^2-\mu_{LLL}^1,  \mu_{DLL}^2 - \mu_{DLL}^1)$ (the green shaded area).
	 Suppose that  the margin of clinical relevance $\Delta_0$ is 0.1, then the acceptance region for null hypothesis $\h_{0,c}$ is the lower  left area of the blue contour line. As there is no intersection between the \sir \ of $(\mu_{LLL}^2-\mu_{LLL}^1,  \mu_{DLL}^2 - \mu_{DLL}^1)$ and the acceptance region for $\hnc$, one may conclude that $\h_{0,c}$ should be rejected.  Alternatively, one can see from the contour lines of $\Delta_{max}$  that the sharp lower bound  for $\Delta_{max}$ is 0.25. As $\Delta_0$ is smaller than $\Delta_{max,slb}$, one also rejects $\h_{0,c}$.
	 
	 \begin{figure}[ht]
	 		 	\vspace{-1cm}
	 	\centering  \includegraphics[width=0.7\textwidth]{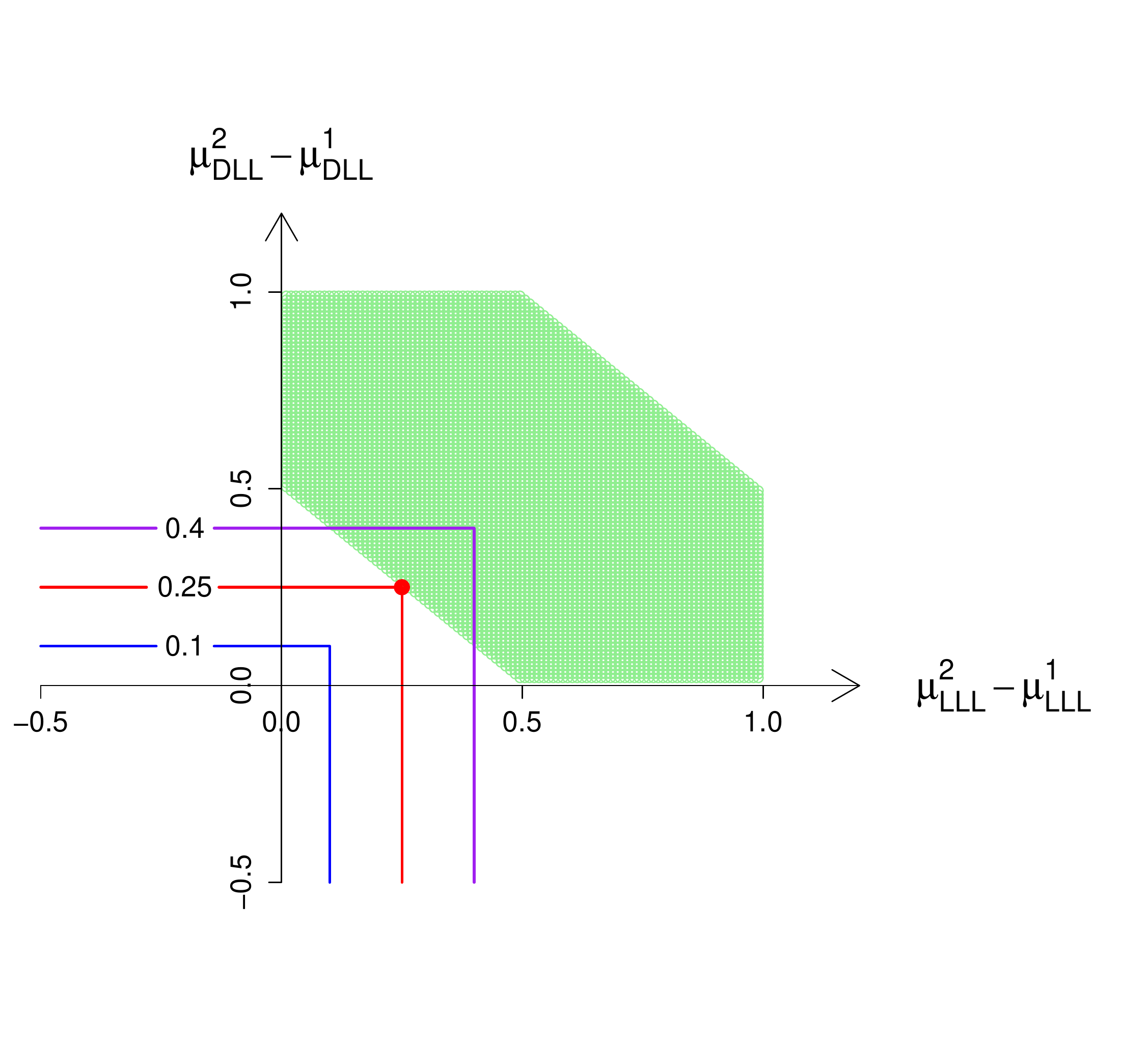}
	 	\vspace{-1cm}
	 	\caption{Feasible region of $(\mu_{LLL}^2-\mu_{LLL}^1, \mu_{DLL}^2-\mu_{DLL}^1)$ (green shaded area).  The colored lines are contour lines of $\Delta_{max}$. The sharp lower bound of $\Delta_{max}$ is obtained at the red point.}
	 	\label{fig:example2}
	 \end{figure}
	 
	 However, by projecting the joint \sir \ of $(\mu_{LLL}^2-\mu_{LLL}^1,  \mu_{DLL}^2 - \mu_{DLL}^1)$ onto individual axises, one concludes that the marginal \sir s  for $\mu_{LLL}^2 - \mu_{LLL}^1$ and $\mu_{DLL}^2-\mu_{DLL}^1$ are both $[0,1]$.  As  both of the marginal \sir s contain values that are smaller than $\Delta_0$,  the data contain no information on the specific contrast that corresponds to the overall treatment effect.   	 
%
%
%
%
	\end{remark}

\section{Data Illustrations}
	\label{sec:data}
	\subsection{Application to the HIV Vaccine Trials Network 503 study}
	
	\label{sec:hvtn}
	
The HIV Vaccine Trials Network (HVTN) 503 HIV vaccine study  was a randomized, double-blinded, placebo-controlled Phase IIb test-of-concept clinical trial to investigate the efficacy and safety of an experimental HIV vaccine. The same vaccine was also evaluated in a different population in an earlier HVTN 502/Step trial. Starting January, 2007, the HVTN 503 study enrolled 800 HIV negative subjects and randomized them to receive three doses of either the study vaccine or a placebo. The ratio of vaccine to placebo assignment was  1:1.  	
	Enrollment and vaccinations were halted in September 2007, but follow-up continued, after the HVTN 502/Step trial met its prespecified non-efficacy criteria.	
	 Details of this study can be found in \cite{gray2011safety,gray2014recombinant}.

	In our analysis, we compared CD4 counts among participants within the same principal stratum defined by their full potential infection statuses.  	
	Due to the early stopping of vaccinations of the trial, a majority of participants in the HVTN 503 trial were not fully immunized. When enrollment was stopped, 400 participants in the HVTN 503 trial were assigned to the experimental vaccine group. Of them, 112  received one injection, 259 received two injections, and only 29 received all three injections.  Hence we considered the dosage of experimental vaccine as the treatment arm $Z$, where $Z=0$ for all subjects in the control group. As the trial was stopped administratively, and the time a participant entered this trial was unlikely to affect the potential outcomes of interest (CD4 count),  it is reasonable to assume that the treatment arms were randomized. Furthermore, since there were only 3.6\% of participants who received all three experimental vaccines, we code $Z=2$ for all participants who receive two or more experimental vaccine injections.

	A total of 100 subjects were infected during this trial.  
 We defined each subject's ``median CD4 count'' (the outcome of interest) as their median CD4 count measured between their confirmatory HIV testing visit and the end of follow-up or start of antiretroviral treatments. 
We also dichotomized CD4 count at 350 cells/mm$^3$ and 200 cells/mm$^3$ as they have been used in previous United States Department of Health and Human Services (DHHS) guidelines for initiating antiretroviral treatment. 
Note that the outcome measure is only measured for infected subjects. As 87.5\% of the study subjects were uninfected, an intent-to-treat analysis with imputation for missing CD4 count values is likely to have very low power for detecting any treatment effects \citep{gilbert2003sensitivity}. Hence SACEs are of interest for analyzing this trial.

		Table 2 in \supple \  summarizes the observed data for the study participants.  {There were 7 infected participants who had no CD4 count measurements after their confirmatory HIV testing visit. We made the missing completely at random (MCAR) assumption and left them out of our analysis below. In treatment arm 0, 1, 2, the mean number of CD4 counts available were 5.69, 5.94 and 5.57, respectively; the mean length of time from the confirmatory HIV testing visit to the first CD4 count measure were 26 days, 25 days and 32 days, respectively, and the mean time spacing between CD4 count measurements were 127 days, 146 days and 134 days, respectively.}

	Presumably there was  little interaction among HVTN 503 subjects so that the SUTVA was plausible. Subsequent analyses of the HVTN 502 and HVTN 503 data suggested that although not possible to directly cause HIV infections itself, the investigational vaccine may increase susceptibility to HIV infection for recipients 
	\citep{gray2011safety,gray2014recombinant}. Given the negative results on the primary efficacy endpoints,  members of  the HVTN 503 Protocol Team  whom we consulted agreed that it is reasonable to make the reverse monotonicity assumption such that experimental vaccine did not help prevent HIV infection for any participant in the study population.  
	The empirical infection rates in the $Z=0,1,2$ arms were $9.25\%, 16.07\%$ and $15.63\%$, respectively.
 Thus, the reverse monotonicity assumption seemed acceptable, and we proceeded with our analysis under this  assumption.

	{
Table \ref{tab:result} summarizes the analysis results. The simultaneous testing method estimates the posterior probability of existence of an overall non-zero treatment effect, while the marginal testing method  estimates the posterior probability that  an overall non-zero treatment effect can be claimed along with the specific treatment arms and principal strata that correspond to this treatment effect. These posterior probabilities were high, suggesting evidence of a non-zero treatment effect on median CD4 falling below 350 or 200 cells/mm$^3$. The 95\% credible intervals for lower bound on $\Delta_{max}$ provide information on the magnitude of vaccine effects. For example, results in Tables \ref{tab:result} show that  there exists at least one basic principal stratum and treatment comparison for which the vaccine  reduces the probability of median CD4 count $\leq$ 200 cells/mm$^3$ by at least 0.026, but we were not able to ascertain the specific basic principal stratum and treatment comparison that corresponds to this effect. The reason for this is two fold. Firstly, because of the  non-identifiability of the SACEs, if the effect size is too small, one may fail to identify the specific causal contrast that corresponds to a clinically relevant treatment effect even with an infinite sample size. Secondly, our proposed methods may deliver more conclusive results if the sample size is large enough. For example, if the sample size was 3000 (which was the estimated sample size in the HVTN 503 trial protocol) and the observed frequencies $P(S=1\mid Z=z)$ and $P(Y=1\mid Z=z,S=1)$ had remained the same, then the 95\% credible interval for the contrast $\mu_{LLL}^2 - \mu_{LLL}^0$ would have been $[0.057, 0.186]$, which would imply that compared to the placebo, receiving  two or more injections of the experimental vaccine is clinically effective for reducing the possibility of very low CD4 cell counts (200 cells/mm$^3$ or less) among subjects who would get infected regardless of which treatment arm they were assigned to.}

{We conclude this part with several caveats. First, the median CD4 count is a non-traditional endpoint for HIV vaccine efficacy trials, and it may not be completely comparable between treatment groups because of differences in the number and timing of CD4 measurements. Second, we have dichotomized CD4 count in our analysis, which results in loss of information.  Third, we have made the MCAR assumption for the missing values in CD4 count measures, which is hard to verify for this data set. Fourth, as pointed out by  some authors \citep[e.g.] []{pearl2011principal},  under the principal stratification framework we have taken here, the vaccine effect estimates are only relevant for the subgroup of subjects who would get infected under at least two dosage levels, which constitutes only a small fraction of the population.
	 Finally, a reduction of 0.026 in the probability of median CD4 counts $\leq$ 200 cells/mm$^3$ may not be considered clinically important given the earlier finding that the vaccine increased HIV acquisition in the study population.}

%
%
%
%

	\begin{table} 
		\begin{center}
			\caption{Posterior probabilities of finding a non-zero overall treatment and posterior credible intervals for lower bounds on $\Delta_{max}$ (the maximal treatment effect over all principal strata and treatment comparisons) for the HVTN 503 trial}
			\begin{tabular}{rrrrrrrrrr}
				\toprule
			   Methods & Posterior probability of a  &  95\% credible interval for \\
			       &  non-zero treatment effect &  lower bound on $\Delta_{max}$  \\
				\midrule
				Outcome:median CD4 $>$ 350 &  &  \\
				Simultaneous &  0.882     &    [0.000, 0.346]    \\
				Marginal &   0.651      &    [0.000, 0.341]        \\
				Outcome:median CD4 $>$ 200 &  &  \\
				Simultaneous &  0.996     &    [0.026, 0.260]    \\
				Marginal &   0.973     &    [$6\times 10^{-4}$, 0.245]        \\				
				\bottomrule
			\end{tabular}
			\label{tab:result}
		\end{center}
	\end{table}

\subsection{Application to survey incentive trials}

Faced with declining voluntary participation rates, there is now a consensus that incentives are effective for motivating response to surveys \citep{singer2002paying,singer2013use}.  There is, however, controversy on how  incentives  affect the quality of data collected. Social exchange theory suggests that by establishing an explicit exchange relationship, incentives not only encourage participation in surveys, but also encourage respondents to provide more accurate and complete information \citep{davern2003prepaid}. However, current experimental studies have mixed findings on this hypothesis \citep{singer2002paying,singer2013use}. 

These experimental studies directly compare response quality in different incentive groups without accounting for the problem of truncation by response. Here the treatments $Z$ are the levels of incentive, the intermediate outcomes $S$ are the responses to the surveys, and the final outcomes $Y$ are measures of survey quality. Although some researchers realize that people persuaded to participate through the use of incentives will have less internal motivation for filling out the survey thoroughly \citep[e.g.][]{davern2003prepaid}, few, if any,  separate this group of people in their analyses from those who would participate in the survey regardless of incentive levels, rendering their results subject to selection bias. Furthermore, arguably the response quality is \emph{undefined} for survey non-respondents. Thus as argued by \cite{rubin2006causal} and others, the naive comparison is not causal as it compares different groups of people at baseline. Instead, for two-arm trials, the SACE is of interest as the subgroup whose members would respond regardless of the level of incentive  is the only group for which both of the potential outcomes are well-defined.  This holds similarly for multiarm trials. Moreover, it is very common that such randomized experiments have multiple incentive groups \citep{singer2002paying,singer2013use}. Hence the methodology introduced in this paper, and more generally, identification and estimation methods for SACEs in multiarm trials are especially relevant.

For example, \cite{curtin2007incentives} used data from the Survey of Consumer Attitudes (SCA) conducted by the University of Michigan Survey Research Center to  investigate whether efforts to increase the response rate jeopardize response quality. Their analysis was based on a random digit dial telephone survey conducted between November 2003 and February 2004. In each of the four months, eligible samples were randomly assigned to one of three experimental conditions: advance letter without an incentive, advance letter plus \$5 incentive and advance letter plus \$10 incentive. The same follow-up procedures, including promised refusal conversion payments are used in all three groups.  The measure for response quality in such studies are inevitably subjective; they can be binary (e.g., ``mostly compete'' vs ``partially complete,'' or whether a particularly important question is answered) or continuous (e.g. percent of missing items). As we don't have access to this data set, below we only discuss the validity of our assumptions.

The SUTVA is reasonable as these are random digit dial samples from the coterminous United States. The monotonicity assumption is also plausible. As argued by survey sampling experts, incentives will motivate response as they  compensate for the relative absence of factors that might otherwise stimulate cooperation \citep{singer2002paying}, so that individuals who would respond with a lower incentive would also respond if offered a higher incentive. Empirical evidence in this study also supports this assumption: the response rates for the three experimental groups were 51.7\%, 63.8\% and 67.7\% \citep{curtin2007incentives}. 

	\section{Discussion}
	\label{sec:discussion}

{	In randomized trials  with truncation by death, the average causal effects in basic principal strata are often of interest as they provide causally meaningful and interpretable  summaries of the treatment effects. However, for trials with multiple treatment arms,  there are usually many such causal contrasts that are of interest to investigators. In this article, we consider testing and estimation problems on the  basic principal stratum causal effects.  Specifically, we propose three scientific questions to understand the overall treatment effect and individual principal stratum causal effects. We then develop novel inference procedures to answer these questions, and show that the proposed procedures have desirable asymptotic properties. }
	
{	Compared to analyzing a multiarm trial in a standard setting, the main difficulty introduced by truncation by death is that the causal estimands are not identifiable. 
	In this case, we show that compared to marginal methods,  the (ANOVA type) simultaneous inference methods  provide more power for testing the overall treatment effect, and the advantage remains even with an infinite sample size. These results demonstrate the importance of addressing both joint and marginal hypotheses in a causal analysis of multiarm trials with truncation by death. This idea may be applied to analyse multiarm trials in other settings in which the causal estimands are not identifiable. For example, in multiarm trials with non-compliance, existing methods consider the causal contrasts separately {\citep{cheng2006bounds,long2010estimating}}.  Although results obtained with such methods are valid, they are often not informative, especially in the case where there are more than three treatment arms {\citep{long2010estimating}}.  In this case, a simultaneous inference method may yield a greater posterior probability of claiming an overall treatment effect and the joint posterior credible intervals are less likely to contain the origin.
}
	
	{
	In analyzing a multiarm trial with truncation by death, researchers may dichotomize the treatment variable to simplify an analysis, especially in settings where the multiarm trials consist of a placebo arm and several dosage groups for an active treatment. One such example is the HVTN 503 study, where the treatment groups 1 and 2 can be considered as different versions of the experimental vaccine. However, as noted by {\cite{hernan2011compound}}, results from  analyses that combine treatment arms in this way may not be generalizable to other population as the causal effect of a compound treatment depends on the distribution of treatment versions in the target population. Moreover, because of the non-identifiability of SACEs, one may fail to find an overall treatment effect that could have been found by applying the proposed simultaneous inference procedure. For example, for the HVTN 503 study, if one were to collapse the active treatment groups into a single compound treatment, then the 95\% credible intervals for the SACE corresponding to this compound treatment would be [0.000, 0.253], with which one could not claim any clinically relevant treatment effect. 
	}

	To account for sampling uncertainty in the observed data distribution, we use Bayesian analysis methods to obtain posterior samples of identifiable quantities $\bm p$. 
	 An alternative Bayesian procedure to our method involves posterior sampling on the mean potential outcomes  $\bm{\mu}_{m-1}$.  This alternative approach would directly yield the posterior rejection rate of $\h_0$ and credible intervals for $\Delta_{max,slb}$ without resorting to techniques we have introduced. However, as $\bm{\mu}_{m-1}$ is not identifiable from the observed data, it turns out that the posterior estimates of $\Delta_{max}$ are extremely sensitive to the prior specification on $\bm{\mu}_{m-1}$. We refer interested readers to \cite{richardson2011transparent} for a further discussion of this issue.

%

		The problem we consider here is similar to an instrumental variable analysis in that both problems can be analysed under the principal stratification framework. When the exposure variable in an instrumental variable analysis is binary, the exclusion restriction assumption  is closely related to the null hypothesis in the truncation by death problem, namely the causal effect  in the always-survivor group is zero.  Hence the approach we develop here may be used to partially test the exclusion restriction assumption of an instrumental variable model.

 There are several possible extensions to our framework. For example,  we have restricted our attention to binary outcomes  in this article. We are currently exploring extensions to deal with continuous and categorical outcomes.  In addition, covariate information may be employed to sharpen bounds on SACEs.  Another possible extension is to introduce sensitivity parameters for better understanding of the causal effects of interest.  The tests and bounds we have developed here correspond to extreme results of corresponding sensitivity analyses.

	\section*{Acknowledgements}
{	
	Research reported in this paper was supported by the National Institute of Allergy and Infectious 
	Diseases (NIAID) of the National Institutes of Health (NIH) under Award Number UM1AI068635. The content is 	
	solely the responsibility of the authors and does not necessarily represent the official views of the NIH or the Department of Veterans Affairs. The authors thank the NIAID-funded HIV Vaccine Trials Network for providing the dataset from the HVTN 503 trial. Furthermore, the   authors thank the participants, investigators, and sponsors of the HVTN 503 trial. 
	 The authors also thank Peter Gilbert and members of  the HVTN Ancillary Study Committee for valuable comments.  Richardson was supported by National Institutes of Health Grant R01 AI032475. 
	 Zhou was supported in
	 part by a US Department of Veterans Affairs, Veterans Affairs Health Administration, Research
	 Career Scientist award (RCS 05-196).
}

	\thispagestyle{empty}
	\bibliographystyle{apalike}
	\bibliography{causal}

	\clearpage
	
	\begin{center}
		
		{\LARGE Supplementary Materials for ``Causal Analysis of Ordinal Treatments and Binary Outcomes under Truncation by Death''} 
		\vspace{1cm}
		
		{\Large Linbo Wang, Thomas S. Richardson  and  Xiao-Hua Zhou} $\ $
		
		
	\end{center}
	\setcounter{equation}{0}
	\setcounter{figure}{0}
	\setcounter{table}{0}
		\setcounter{algorithm}{0}
	\setcounter{page}{1}
	\makeatletter
	\renewcommand{\theequation}{S\arabic{equation}}
	\renewcommand{\thefigure}{S\arabic{figure}}
		\renewcommand{\thetable}{S\arabic{table}}
				\renewcommand{\thealgorithm}{S\arabic{algorithm}}
	\setcounter{section}{0}

		\section{Algorithm for identifying $\Delta_{max,slb}$}
		
		See Algorithm \ref{alg:bound_global_binary}. 
		
		\begin{algorithm}
			\caption{\ \ An algorithm for identifying  $\Delta_{max,slb}$}
			\label{alg:bound_global_binary}
			\begin{enumerate}
				\item Solve the following linear programming problem:\\
				minimize $\alpha$  subject to:
				\begin{flalign*}
				\sum\limits_{g\in \Omega_z} p_{g}^z  \mu_{g}^z  = m(z), &\quad z = 0,\ldots,m-1; \\
				\max\left(0, m(z) -  p_{D^m L}^z \right) \leq \sum\limits_{g\in \Omega_{m-1}} p_{g}^z \mu_g^z \leq \min \left( 1-    p_{D^m L}^z,      m(z)    \right),  &\quad z =  m; \\
				\mu_{g}^{z_1} - \mu_{g}^{z_2} \leq \alpha, &\quad \forall g,  z_1 \geq z_2; \\
				0 \leq \mu_g^z \leq 1, &\quad \forall g, z
				\end{flalign*}
				\item Report the value of the linear programming problem above as $\Delta_{max,slb}$
			\end{enumerate}
		\end{algorithm}

		\section{Simulation studies}
		\label{sec:simu}
		
		We now use a hypothetical example to illustrate the  advantage of the simultaneous inference procedures proposed in Section 3 and 4 in the main text for testing the overall treatment effect. Let  the comparison method be the approach that considers each $\Delta(z_1,z_2;g)$ separately, and it accepts or rejects the null  based on the marginal \sir s of $\Delta(z_1,z_2;g)$. With the comparison marginal testing method, one rejects the hypothesis $\h_0$ only if at least one of the marginal feasible regions excludes 0. In other words, the comparison method rejects $\h_0$  if the observed data not only provide evidence for existence of a non-zero treatment effect, but also contain information on the specific principal strata and treatment arms that correspond to this treatment effect.  As explained in Remark 3 in the main text, this generally yields a smaller  posterior rejection probability. In addition, with the comparison method, one estimates the lower bound on $\Delta_{max}$ to be the maximal sharp lower bound for all $\Delta(z_1,z_2;g)$ that appear in equation (1) in the main text.  We denote this lower bound as $\Delta_{max,mlb}$, where ``mlb'' is short for ``marginal lower bound.'' One can see from the numerical example in Remark 3 in the main text that $\Delta_{max,mlb}$ is in general no larger than  $\Delta_{max}$. This is because the comparison marginal estimation method does not use information on the dependence among feasible regions of  causal contrasts $\Delta(z_1,z_2;g)$. In the simulation studies,  we empirically evaluate  the difference between the proposed simultaneous inference methods and the comparison marginal inference methods for testing the overall treatment effect.

		Suppose that we have a three-arm vaccine trial with two vaccine groups and one placebo group, and there are 400 subjects in each group. The hypothetical data example is listed in Table \ref{tab:hyp_data}, where $n_1$ is a parameter taking integer values between 0 and 40. The conditional frequencies $m(0)$, $m(1)$  and $m(2)$ in this example are $0.025n_1$, 0.7 and 0.9, respectively. In our example, there are 10\% of the study sample in each of the principal strata $LLL, DLL, DDL$, while the rest belongs to the $DDD$ stratum.
		
		\begin{table} 
			\begin{center}
				\caption{Observed data counts in a hypothetical example.}			\begin{tabular}{rrrrrrrrrr}
					\toprule
					Observed subgroup & Counts  \\
					\midrule
					$Y=1,S=1,Z=0$ &  $n_1$        \\
					$Y=0,S=1,Z=0$ & $40-n_1$         \\
					$S=0,Z=0$       &  360       \\ \\ [-5pt]
					$Y=1,S=1,Z=1$ &  56       \\
					$Y=0,S=1,Z=1$ &     24     \\
					$S=0,Z=1$		&	320		\\ 
					\\[-5pt]
					$Y=1,S=1,Z=2$ &  108     \\
					$Y=0,S=1,Z=2$ &   12      \\
					$S=0,Z=2$		&  280	   \\
					\bottomrule
				\end{tabular}
				\label{tab:hyp_data}
			\end{center}
		\end{table}

		Results in Figure \ref{fig:simu} show  that for some values of $n_1$, the simultaneous and marginal methods compared here yielded similar results. However, in some other cases, the results could be very different. For example, when $n_1=36$, the simultaneous testing method estimated  the posterior probability of rejecting $\h_0$ to be 98.8\%, compared to an estimate of 4.0\% from the marginal testing method. When $n_1=20$, the simultaneous estimation method estimated the 95\% credible interval for $\Delta_{max,slb}$ to be $[0.029,0.404]$, based on which one was able to claim a clinically relevant treatment effect at margin $\Delta_0=0.02$. The marginal estimation method, however, estimated   the 95\% credible interval for  $\Delta_{max,mlb}$ to be $[4\times 10^{-4},0.363]$, with which one  failed to claim a clinically relevant treatment effect at the same margin.

		\begin{figure}
			\begin{minipage}[t]{0.5\linewidth}
				\centering
				\includegraphics[width=2.5in]{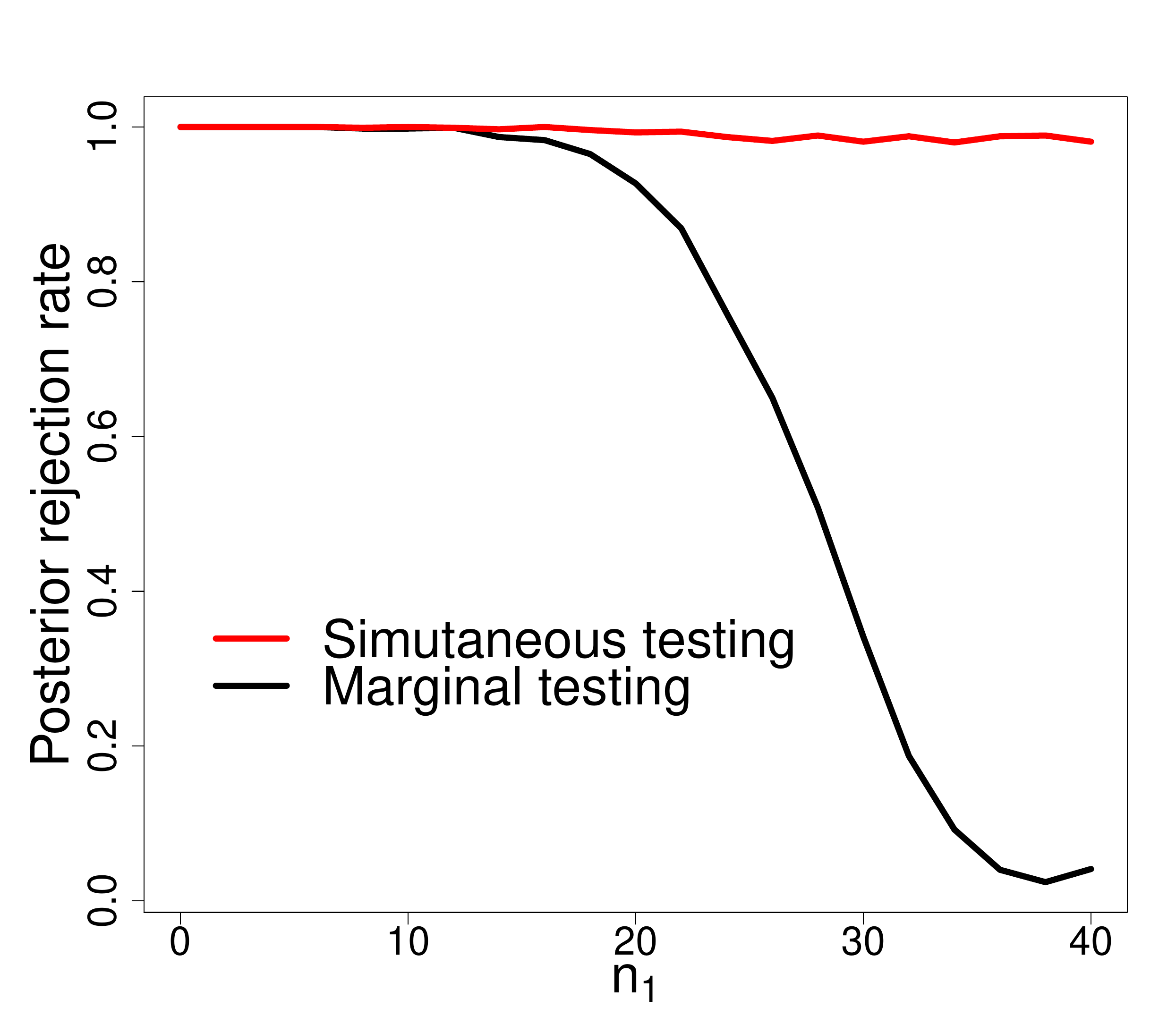}	    
			\end{minipage}%
			\begin{minipage}[t]{0.5\linewidth}
				\centering
				\includegraphics[width=2.5in]{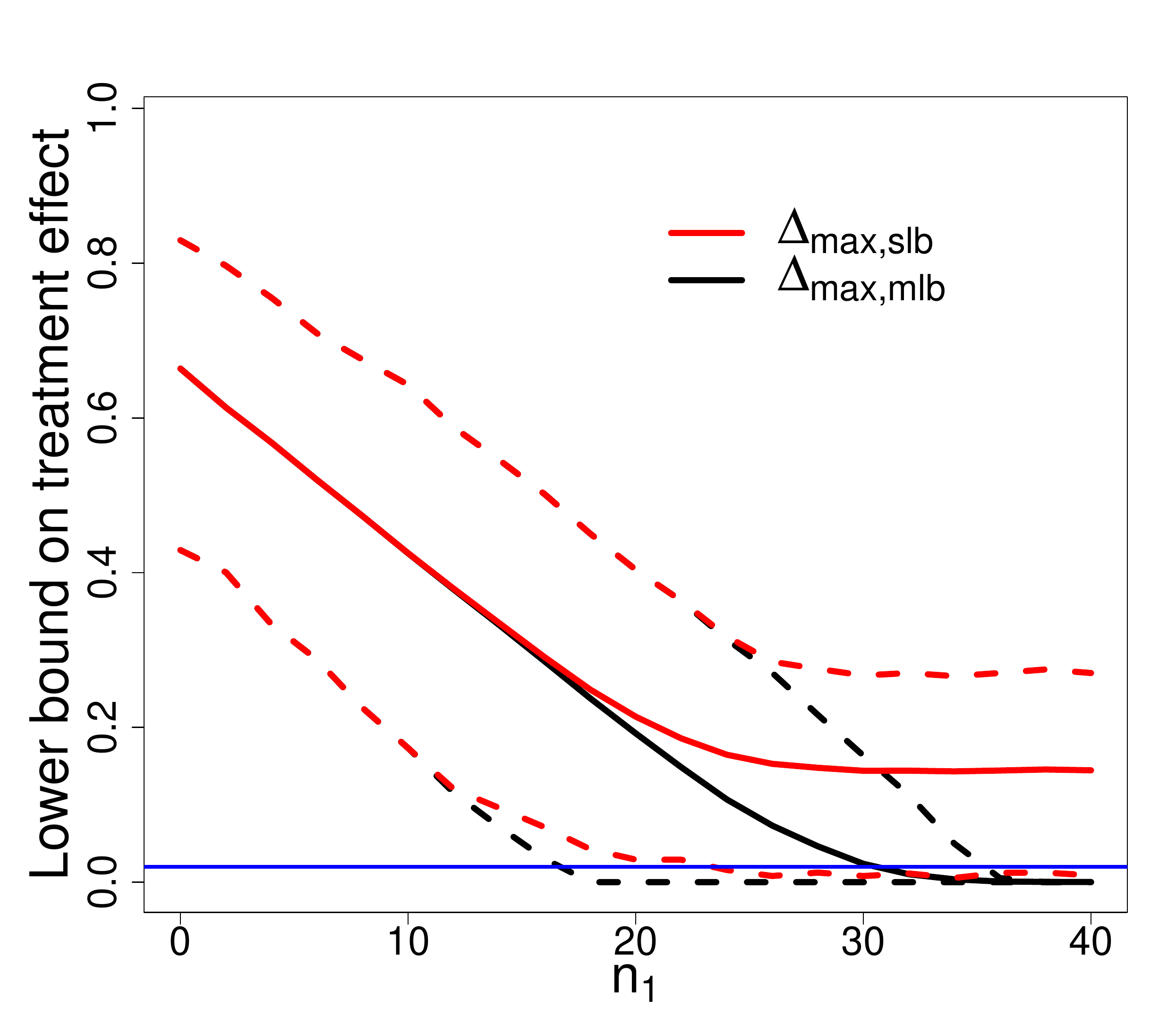}
			\end{minipage}
			\caption{Results from analyzing the hypothetical data set in Table \ref{tab:hyp_data}. The left panel shows the posterior probability of rejecting $\h_0$ using the proposed simultaneous testing method and the comparison marginal testing method. The right panel shows the posterior mean (solid lines) and 95\% credible intervals (dashed lines) for lower bounds on $\Delta_{max}$, the maximal treatment effect among all possible basic principal strata and treatment comparisons. The red curves correspond to sharp lower bounds obtained using the proposed simultaneous estimation method, and the black curves correspond to lower bounds obtained using the comparison marginal estimation method. The blue horizontal line corresponds to a clinically meaningful margin of 0.02.}
			\label{fig:simu}
		\end{figure}

		\section{Data Table for the HVTN 503 study}
		
		Table \ref{tab:obs_data} gives the observed data counts for the HVTN 503 trial.
		
		\begin{table} 
			\begin{center}
				\caption{Observed data counts in the HVTN 503 trial. $Z$ denotes the treatment arm, $S$ denotes the infection status, and $Y$ is the dichotomized outcome of CD4 count. $Y=*$ denotes that $Y$ is missing.}
				\begin{tabular}{rrrrrrrrrr}
					\toprule
					Observed subgroup & median CD4 $>$ 350 cells/mm$^3$ & median CD4 $>$ 200 cells/mm$^3$  \\
					\midrule
					$Y=1,S=1,Z=0$ &  19    &  29    \\
					$Y=0,S=1,Z=0$ & 14      &   4   \\
					$Y=*,S=1,Z=0$ & 4      &   4   \\
					$S=0,Z=0$       &  363     &  363     \\ \\ [-5pt]
					$Y=1,S=1,Z=1$ &  12     &   16   \\
					$Y=0,S=1,Z=1$ &    4  &   0  \\
					$Y=*,S=1,Z=1$ &    2  &   2   \\
					$S=0,Z=1$		&	94	& 	94	\\ 
					\\[-5pt]
					$Y=1,S=1,Z=2$ &  34     &   44   \\
					$Y=0,S=1,Z=2$ &   10    &    0  \\
					$Y=*,S=1,Z=2$ &   1    &    1  \\
					$S=0,Z=2$		& 243	&  243     \\
					\bottomrule
				\end{tabular}
				\label{tab:obs_data}
			\end{center}
		\end{table}

		\section{Proofs of theorems and lemmas}
		\renewcommand{\thesubsection}{\Alph{subsection}}

		\subsection{Proof of Theorem 1}
		
		The proof for the general multi-arm case is very similar to the discussion for the three-arm case. The only non-trivial generalization is for Step 3 of Algorithm 1 in the main text. Instead of checking the pairwise intersections of $(B_{kz}; z=k,\ldots,m)$, we check their joint intersection. This relies on the observation that if we let $g=D^k L^{m+1-k}$, then $\omega_g^k = 1$ and  the feasible region for $B_{kk}$ is a one point set $\{\int y dQ_{g}^k\}$.
		Consequently, 
		\begin{equation}
		\label{eqn:inter_joint}
		\mathop{\cap}\limits_{z=k,\ldots,m} B_{kz} \neq \emptyset
		\end{equation}
		implies  that
		\begin{equation}
		\label{eqn:inter_pariwise}
		B_{kz_1} \cap B_{kz_2} \neq \emptyset, \forall z_1 > z_2 \geq k.
		\end{equation}
		Note there are only $m-k$ pairs of comparisons involved in \eqref{eqn:inter_joint}, compared to
		$(m+1-k)(m-k)/2$ pairs of comparisons in \eqref{eqn:inter_pariwise}.

		\subsection{Proof of Theorem 2}
		\label{appen:proof}

		To prove Theorem 2,  we note that the assumptions of Theorem 2 and the  \odd\  impose the following constraints on 
		$\mu_g^z$:
		\begin{flalign}
		Q^z &= \sum\limits_{\ug \in \Omega_{l-1}}  p_{\ug}^z Q_{\ug}^z + p_g^z Q_g^z + \sum\limits_{\upg \in \Omega_z \setminus \Omega_{l}}  p_{\upg}^z Q_{\upg}^z \label{eqn:contrainsts_odd},  \\
		\mu_{\ug}^{\mathcal{M}(\ug)} &= \cdots = \mu_{\ug}^{m}. \forall \ug \in \Omega_{l-1}, \numberthis\label{eqn:constrainsts_working} 
		\end{flalign}
		where $Q^z$ denotes the distribution of  outcome $Y$ in treatment arm $z$. 
		To simplify \eqref{eqn:contrainsts_odd} and \eqref{eqn:constrainsts_working}, we  use the following lemmas, which say that both the proportions of basic principal strata $p_g^z$ and the means of Bernoulli distributions $(Q_{\ug}^z, \ug\in \Omega_{l-1}, z\geq \mathcal{M}(\ug))$ are identifiable. Proofs of these lemmas are left to the end of this subsection.  
		
		\begin{lemma}
			\label{lemma:pgz}
			The proportions of basic principal strata, namely $(p_g^z; g\in \Omega_{m-1}, z \geq \mathcal{M}(g))$ are identifiable from the observed data. 
		\end{lemma}

		\begin{lemma}
			\label{lemma:mugz}
			Suppose that (6) in the main text holds for all $k < l$, then $(\mu_{\ug}^z; \ug\in \Omega_{l-1}, z\geq \mathcal{M}(g))$ are identifiable from the observed data.
		\end{lemma}
		%
		%
		As the  Bernoulli distribution $Q_{\ug}^z$ is uniquely determined by its mean $\mu_{\ug}^z$,  the constraints on $\mu_g^z$ can be simplified as 
		\begin{equation}
		\label{eqn:constraints_simplified}
		Q_{\bm g}^z = 
		\omega_{g}^z Q_{g}^z + \sum\limits_{\upg \in \Omega_z \setminus \Omega_{l}}  \omega_{\upg}^z  Q_{\upg}^z,
		\end{equation}
		where $Q_{\bm g}^z$  a Bernoulli distribution with mean $m_{\bm g}(z)$. Applying 	 \cite{imai2008sharp}'s results to \eqref{eqn:constraints_simplified}, we have  \begin{equation*}
		B_{lz} = \left[\int y d L_{\omega_{g}^z}(Q_{\bg}^z),         \int y d U_{\omega_g^z}(Q_{\bg}^z)\right].
		\end{equation*}
		This completes the proof of Theorem 2.
		\qed
		
		\bigskip
		\noindent{\bf{Proof of Lemma \ref{lemma:pgz}}}
		\begin{proof}
			Let $\pi_g^z = P(G=g|Z=z)$. 
			Following Assumption 2, $\pi_g^z$ is independent of treatment arm $z$ and hence can be written as $\pi_g$. Under Assumption 3, we have the following equations:
			\begin{flalign*}
			P(S=1|Z=0) &= \pi_{L^{m+1}}, \\
			P(S=1|Z=1) &= \pi_{L^{m+1}} + \pi_{DL^{m}}, \\
			\quad & \cdots \\
			P(S=1|Z=z) &= \pi_{L^{m+1}} + \ldots + \pi_{D^z L^{m+1-z}}, \numberthis\label{eqn:pi_iden}  \\
			\quad  & \cdots \\
			P(S=1|Z=m) &= \pi_{L^{m+1}} + \ldots + \pi_{D^m L},\\
			1 &=  \pi_{L^{m+1}} + \ldots + \pi_{D_{m+1}}.
			\end{flalign*}
			It can be shown that there exists an unique solution to equation (\ref{eqn:pi_iden}) and hence
			($\pi_g, g\in\Omega_m$)  are identifiable from equation \eqref{eqn:pi_iden}. It then follows that $(p_g^z; g\in \Omega_{m-1}, z \geq \mathcal{M}(g))$ are also identifiable.
		\end{proof}
		
		\bigskip
		\noindent{\bf{Proof of Lemma \ref{lemma:mugz}}}
		\begin{proof}
			As (6) in the main text holds for all $k < l$, 
			we only need to show that $\mu_{\ug}^{\mathcal{M}(\ug)}$ is identifiable from the observed data. We show this by applying  the induction method on $\mathcal{M}(\ug)$.
			
			Base case: if $\mathcal{M}(\ug) = 0$, then $\mu_{\ug}^{\mathcal{M}(\ug)} = \mu_{L^{m+1}}^0=P(Y=1|Z=0,S=1)$ by the monotonicity assumption (Assumption 3). 
			
			Inductive step: suppose that $\mu_{\ug}^{\mathcal{M}(\ug)}$ is identifiable from the observed data for all principle strata $\ug$ such that $\mathcal{M}(\ug) \leq k$. Following  the monotonicity assumption (Assumption 3), we have the following identify:
			\begin{flalign*}
			P(Y=1|Z=k+1,S=1) &= \sum\limits_{\ug \in \Omega_{k}} p_{\ug}^k \mu_{\ug}^k	+ p_{D^{k+1} L^{m-k}}^{k+1} \mu_{D^{k+1} L^{m-k}}^{k+1} \\
			&= \sum\limits_{\ug \in \Omega_{k}} p_{\ug}^k \mu_{\ug}^{\mathcal{M}(\ug)}	+ p_{D^{k+1} L^{m-k}}^{k+1} \mu_{D^{k+1} L^{m-k}}^{k+1}\numberthis\label{eqn:lemma2_3},
			\end{flalign*}
			where the last step in \eqref{eqn:lemma2_3} follows from the working hypotheses.
			
			Following Lemma \ref{lemma:pgz}, $(p_{\ug}^k; \ug \in \Omega_k)$ and $p_{D^{k+1} L^{m-k}}^{k+1}$ are identifiable from the observed data. Following the induction hypotheses, $(\mu_{\ug}^{\mathcal{M}(g)}; \ug \in \Omega_k)$ are also identifiable. Consequently, $\mu_{D^{k+1} L^{m-k}}^{k+1}$  is identifiable from \eqref{eqn:lemma2_3}.  In other words, for principle strata $\ug$ such that $\mathcal{M}(\ug) = k+1$, $\mu_{\ug}^{\mathcal{M}(\ug)}$ is also identifiable from the observed data.
			
			By the induction principle, we have finished our proof.
			%
		\end{proof}

		\subsection{Proof of Theorem 4}
		
		Theorem 4 is a direct consequence of the following lemma:
		
		\begin{lemma}
			\label{lemma:lemma1}
			Let $h$ be a mixture of $k$ Bernoulli distributions $f_1, \ldots, f_k$: $h = \sum\limits_{j=1}^k \alpha_j f_j$, where the mixing proportions $\alpha_j, j=1,\ldots,k$  are known. Let $P, P_1, \ldots, P_k$ be the probability of a positive outcome under $h, f_1,\ldots, f_k$ respectively, then
			$$
			\max\left(0, P-\sum\limits_{j=l+1}^k \alpha_j \right)    \leq       \sum\limits_{j=1}^{l} \alpha_j P_j \leq
			\min\left(\sum\limits_{j=1}^l \alpha_j, P \right).
			$$
		\end{lemma}
		Lemma \ref{lemma:lemma1} is a generalization of Lemma 1 in \cite{cheng2006bounds} and can be proved by solving the linear programming problem of minimizing or maximizing $ \sum\limits_{j=1}^{l} \alpha_j f_j $ subject to constraints $P = \sum\limits_{j=1}^k \alpha_j P_j$.   \qed

\end{document}